\documentclass[aps, showpacs, pra, superscriptadaddress, 
twocolumn] {revtex4-1}
\usepackage{hyperref}
 \usepackage{amsmath}
\usepackage{amssymb}
\usepackage{epsfig}
\usepackage{natbib}
\usepackage{epstopdf}
\usepackage{graphicx}

\usepackage{floatrow}
\usepackage[caption=false,font=Large]{subfig}

\newcommand*{\be}{\begin{equation}}
\newcommand*{\ee}{\end{equation}}
\newcommand*{\bea}{\begin{eqnarray}}
\newcommand*{\eea}{\end{eqnarray}}

 \DeclareFontFamily{OT1}{pzc}{}
 \DeclareFontShape{OT1}{pzc}{m}{it}%
 {<->  s  *  [1.400]  pzcmi7t}{}
\DeclareMathAlphabet{\mathscr}{OT1}{pzc}%
{m}{it}

\begin{document}

\title{
An exactly solvable $\mathcal{PT}$-symmetric  dimer  \\
from a  Hamiltonian system  of nonlinear 
oscillators with gain and loss}

\author{I V Barashenkov}    
\affiliation{
 Department of Mathematics and Centre for Theoretical and Mathematical Physics,  University of Cape Town, 
 Rondebosch 7701, South Africa}

 \author{Mariagiovanna Gianfreda}
\affiliation{ Department of Physics, Washington University, St. Louis,
MO 63130, USA }

\begin{abstract}
We show that a pair of coupled nonlinear oscillators, of which
one oscillator has positive and the other one negative damping of equal rate, can form a Hamiltonian system.
Small-amplitude oscillations in this  system are governed by a  $\mathcal{PT}$-symmetric nonlinear Schr\"odinger dimer
with linear and cubic coupling. 
The  dimer also represents a Hamiltonian system and
is found to be exactly solvable in elementary functions.
We show that the nonlinearity softens
the $\mathcal{PT}$-symmetry breaking transition
in the nonlinearly-coupled dimer:
stable periodic and quasiperiodic  states
with large enough amplitudes persist for an arbitrarily large value of the gain-loss coefficient.
\end{abstract}

\pacs{}
\maketitle

\section{Introduction}

Originally introduced as a concept in quantum mechanics
\cite{Bender}, the idea of $\mathcal{PT}$ symmetry has expanded into a wide range of fundamental and 
applied sciences,
most notably  photonics \cite{optics,Kip,optics_2},  plasmonics \cite{Lupu}, quantum optics of atomic gases \cite{atomic},
studies of the Bose-Einstein condensation \cite{BEC,Graefe},
and physics of electronic circuits \cite{electronics}.
The $\mathcal{PT}$-symmetric equations 
model physical structures with  balanced gain and loss.
These lie halfway between open systems (systems in contact with 
 external environment) and closed, isolated, settings.
Increasing the gain-loss rate takes the structure from
the unbroken $\mathcal{PT}$-symmetry phase,  
 characterised by stationary, periodic or quasiperiodic evolution, 
 to the symmetry-broken phase,  where  
 it is in the uncontrollable blow-up regime.

The  transition between the two phases  has 
been observed in a variety of  experimental environments  including 
optics \cite{optics,Kip},
superconductivity 
\cite{exp1},  microwave cavities \cite{exp2}, atomic diffusion 
\cite{exp3}, and nuclear magnetic resonance \cite{exp4}. In particular, the $\mathcal{PT}$-symmetric  phase 
transition is easily  recognised in structures consisting of
coupled  oscillators with gain and loss  \cite{electronics,exp6,exp7,TL}.
A theoretical modelling of this experimental setting \cite{exp7} has demonstrated that
 two coupled linear oscillators with balanced gain and loss
can form  a Hamiltonian system \cite{BG}.

The system considered in \cite{BG} (see also \cite{electronics}) had the form
\begin{align}
{\ddot x} + 2 \eta {\dot x}
+ x + 2 \kappa y =0, \nonumber
 \\
{\ddot y} - 2 \eta {\dot y} 
+ y + 2 \kappa x=0.
\label{A1}
\end{align}
Here $x$ and $y$ are the coordinates of  two coupled harmonic 
oscillators --- or the two degrees of freedom of a particle in a parabolic well. The coefficient
 $\eta>0$  gives the rate of damping experienced 
by the $x$-component and, at the same time, 
quantifies the energy gain by the component $y$.
The coefficient $\kappa>0$ measures the coupling between the two components.
Finally, the overdot stands for the derivative in $t$.

Given the fact that there are channels both for the gain and loss of energy,
the availability of the  Hamiltonian for \eqref{A1} is surprising and counter-intuitive.
A natural question, therefore,  is how structurally stable this property is.
Will the Hamiltonian structure survive the addition of nonlinear terms?
If yes, how different are dynamical and bifurcation properties of the 
Hamiltonian 
$\mathcal{PT}$-symmetric system from those of its non-Hamiltonian 
counterparts?

The first of these two questions is answered here by 
devising  a Hamiltonian system of two
coupled cubic oscillators whose linear truncation is  given by  Eq.\eqref{A1}.
Assuming that
the gain-loss coefficient  is  small while the coupling is weak,  
we show that the amplitude of the
small-amplitude oscillations in this system
satisfies a two-site discrete nonlinear Schr\"odinger equation with gain in one site and loss in the other.
Like the parent system of two coupled
oscillators, this $\mathcal{PT}$-symmetric Schr\"odinger dimer is Hamiltonian; furthermore, the 
dimer is found to be exactly solvable in elementary functions.
The bulk of our paper is concerned with the 
 analysis of this new discrete nonlinear Schr\"odinger equation.
 
 We compare the Hamiltonian $\mathcal{PT}$-symmetric dimer ---  which 
 features a {\it nonlinear} coupling of the two sites in addition to the standard, linear, coupling ---
 to the previously considered {\it linearly}-coupled $\mathcal{PT}$-symmetric  Schr\"odinger model. 
 One of the striking differences between the two systems is 
 that all stationary states in the new dimer are stable. 
 The other one is that the nonlinearity softens
  the $\mathcal{PT}$-symmetric phase transition
  in the new model.
By this,   we mean that the increase of the gain-loss coefficient beyond the 
point of the linear $\mathcal{PT}$-phase transition does 
not eliminate 
stable stationary and periodic states with large amplitudes.
Stable bounded solutions
persist for an
arbitrarily large value of the gain-loss coefficient.

The outline of this paper is as follows.
In section \ref{Sec2} we introduce our Hamiltonian system of two
coupled anharmonic oscillators with positive and negative damping. 
Subsequently (section
\ref{Sec3}) the two-oscillator system is reduced to a $\mathcal{PT}$-symmetric
discrete Schr\"odinger equation for the amplitudes of  small 
$x$- and $y$-oscillations.
In section \ref{Sec4} we obtain, explicitly, the general solution of 
this Schr\"odinger dimer and in section \ref{Sec5} 
classify its most important, stationary, regimes.
(Technical details of the fixed-point analysis have been relegated to two 
appendices.)
The  bifurcation diagram for the stationary regimes  is compared to the 
corresponding diagram for the linearly-coupled dimer
(section \ref{Discussion}). 
Finally, section \ref{Summary} summarises results of this project.

\section{Hamiltonian system of two oscillators with gain and loss}
\label{Sec2}

The system we propose as a nonlinear extension of
Eqs.\eqref{A1}, is
\begin{align}
{\ddot x} + 2 \eta {\dot x}
+ x + 2 \kappa y + x(x^2+ 3y^2)=0, 
\nonumber
 \\
{\ddot y} - 2 \eta {\dot y}
 + y + 2 \kappa x+ y(y^2+ 3x^2)=0.
\label{soft}
\end{align}
The system \eqref{soft} is $\mathcal{PT}$-symmetric, that is, invariant under the 
joint action of the $\mathcal P$ and $\mathcal T$ transformations.
Here $\mathcal P$ is the operator that swaps the $x$ and $y$ components,
and $\mathcal{T}$ is the time inversion:  $\mathcal T x(t)=x(-t)$, $\mathcal T y(t)=y(-t)$.

One can readily check that the cubic system \eqref{soft} is also Hamiltonian,
with the Hamilton function
\begin{align*}
H= pq - \eta (xp-yq)
+(1-\eta^2) xy +\kappa (x^2+y^2)
\nonumber   \\
+ xy^3 +x^3y.
\label{C1}
\end{align*}
One pair of the Hamilton equations is
\[
{\dot x}= \frac{\partial H}{\partial p}= q-\eta x,
\quad
{\dot y}= \frac{\partial H}{\partial q}= p+ \eta y.
\]
These express the canonical momenta in terms of  the
velocities:
\be
q= {\dot x} + \eta x, \quad p= {\dot y} - \eta y.
\label{C2}
\ee
The second pair is
\be
{\dot p}= -\frac{\partial H}{\partial x},
\quad
{\dot q}= -\frac{\partial H}{\partial y}.
\label{C20}
\ee
Substituting \eqref{C2} in 
\eqref{C20} gives the system \eqref{soft}. 

There is an extensive literature on {\it quantum\/} $\mathcal{PT}$-symmetric Hamiltonians
(see e.g. \cite{QM}).
In particular,  a well established  fact is that a $\mathcal{PT}$-symmetric Hamiltonian operator
can be  related to an equivalent isospectral Hermitian Hamiltonian by a similarity
transformation \cite{Mostafa}.
At the {\it classical} level,
the studies of $\mathcal{PT}$-symmetric Hamiltonian systems were confined to formal 
aspects such as trajectories on the complex plane
  \cite{Bender_class}.
So far, the only example of a real classical $\mathcal{PT}$-symmetric 
system 
with loss and gain that admits a Hamiltonian formulation,  was the set of linear equations \eqref{A1}.
Equations \eqref{soft} constitute the first example of a nonlinear 
system of that kind.

\section{$\mathcal{PT}$-symmetric dimer}
\label{Sec3}

Linearising Eqs.\eqref{soft} about the trivial fixed point 
$x=y=0$, the stability eigenvalues are found to be
\[
(\lambda^2)_{1,2}=2 \eta^2-1 \pm 2 \sqrt{\kappa^2-\eta^2(1-\eta^2)}.
\]
The fixed point is stable  if 
both values for $\lambda^2$ are real and nonpositive; this happens if
the following two conditions are met simultaneously:
\be
\label{A501}
\eta^2 \leq \frac12, 
\quad 
\eta^2(1-\eta^2) \leq \kappa^2  \leq \frac14.
\ee

In this study, we restrict ourselves to small $\eta$ and $\kappa$.
According to Eq.\eqref{A501}, there is a subregion of the small-parameter domain
where 
the fixed point is stable (a centre).
With $\eta$ and $\kappa$ chosen in the   stability subregion,
we expect to find  bounded motions 
in a neighbourhood of the fixed point.
 To construct these quasiperiodic orbits, and
examine their stability, we use the multiple scale expansion.

Letting 
\be
2 \kappa= {K} \epsilon^2, \quad
2 \eta= \Gamma \epsilon^2,
\label{A6}
\ee
we expand $x$ and $y$ in odd powers of $\epsilon$:
\be
x= \epsilon x_1 + \epsilon^3 x_3 + ..., 
\quad
y= \epsilon y_1 + \epsilon^3 y_3 +...
\label{A7}
\ee
Assuming that $x_i$ and $y_i$ depend on a hierarchy of time scales,
$T_0=t$, $T_{2n}=\epsilon^{2n} t$  ($n=1,2,...$),
 gives
\be
\frac{d}{dt}= D_0+ \epsilon^2 D_2 + ...;
\quad
\frac{d^2}{dt^2}= D_0^2 + 2 \epsilon^2 D_0 D_2 + ...
\label{A8}
\ee
Substituting \eqref{A6}-\eqref{A8} in 
\eqref{soft}, we equate coefficients of like powers of $\epsilon$.

The order $\epsilon^1$ produces
\[
(D_0^2+1) x_1= (D_0^2+1) y_1=0,
\]
whence
\be
x_1= {\mathcal A} e^{i T_0} + c.c., \quad
y_1= {\mathcal B}  e^{i T_0} + c.c..
\label{A9}
\ee
Here ${\mathcal A}={\mathcal A} (T_2, T_4, ...)$, ${\mathcal B}={\mathcal B}(T_2, T_4, ...)$,
and $c.c.$ stands for the complex conjugate of the preceding term.
At  the order $\epsilon^3$ we obtain
\begin{align*}
(D_0^2+1) x_3+ (2D_0 D_2 + \Gamma D_0) x_1 + Ky_1+ x_1^3+3y_1^2 x_1=0,   
\\
(D_0^2+1) y_3 + (2D_0 D_2 - \Gamma D_0) y_1+Kx_1 + y_1^3+ 3x_1^2y_1=0.
\end{align*}
Substituting for $x_1$ and $y_1$ from \eqref{A9}, 
and setting the secular terms to zero results in
\begin{align*}
2i D_2 {\mathcal A} + i\Gamma {\mathcal A} + K {\mathcal B} + 3 (|{\mathcal A}|^2+ 2|{\mathcal B}|^2)
{\mathcal A} +3 {\mathcal B}^2 {\mathcal A}^*
=0,   \\
2i D_2 {\mathcal B} - i\Gamma {\mathcal B}+ K{\mathcal A}  + 3 (2|{\mathcal A}|^2+ |{\mathcal B}|^2)
{\mathcal B} +3 {\mathcal A}^2 {\mathcal B}^*
=0.
\end{align*}
Letting  $\tau=K T_2/2$, defining $\gamma =\Gamma/K$, 
and scaling the amplitude components as ${\mathcal A}= (K/3)^{1/2} \psi_1$ and ${\mathcal B}=(K/3)^{1/2}\psi_2$, these equations
 acquire the form 
\begin{align}
i {\dot  \psi_1}   +  \psi_2
+  (|\psi_1|^2+ 2 |\psi_2|^2) \psi_1   
+  \psi_2^2 \psi_1^* &   =    -      i   \gamma \psi_1,     \nonumber 
\\
i {\dot \psi_2}  +  \psi_1 
+  (|\psi_2|^2+2|\psi_1|^2) \psi_2  
+ \psi_1^2 \psi_2^*  &  =   \phantom{-}   i    \gamma \psi_2.  
\label{sys}
\end{align}
(Here and below  the overdot is used to denote the derivative with respect to $\tau$.)

The system \eqref{sys} is in the form of  a two-site discrete nonlinear 
Schr\"odinger equation, the so-called nonlinear Schr\"odinger dimer.
Dimers with various nonlinearities are workhorses of  photonics, where they serve to model stationary light beams 
 in  coupled optical waveguides \cite{couplers,Kip,Dimer_integrability,SXK}.
 (Similar equations govern electromagnetic waves with orthogonal  polarisations 
 propagating in a
single-mode nonlinear fiber, see e.g. \cite{Added_Boris}.)
 The coupler described by \eqref{sys} consists of a waveguide with loss and a guide with an equal amount of optical gain.
The variables
$\psi_1$ and $\psi_2$ represent the corresponding complex beam amplitudes,  $\gamma>0$ is their common gain-loss rate,
and $\tau$ measures the distance along the parallel cores.  
The quantities $P_2=|\psi_2|^2$
and $P_1=|\psi_1|^2$ give
the powers carried by the active and lossy channel,
respectively.

Another area where 
the nonlinear Schr\"odinger dimers occur commonly,  comprises
 the  studies of the boson condensation \cite{Rag,Graefe}. 
In particular, 
the nonlinearity \eqref{sys} describes the mean-field condensate
wave function in a
symmetric double-well potential 
in the two-mode approximation
\cite{Theocharis}.
The $\psi_1$ and $\psi_2$ are the complex amplitudes of the ground 
and the first excited state, respectively.
In the matter-wave context, $P_1$ and $P_2$ are 
the numbers of particles associated with the two modes.

The dynamical regimes in Eqs.\eqref{sys}  are selected by 
varying the gain-loss rate, $\gamma$.
This is a single parameter in the system.
We note that  $\gamma$ admits a
simple expression in terms of the 
 parameters of the original two-oscillator model \eqref{soft}:
$\gamma= \eta/\kappa$.


As its parent system \eqref{soft}, the dimer
 \eqref{sys} is $\mathcal{PT}$-symmetric.
Here the $\mathcal P$ operator is defined by
\be
{\mathcal P} \left(
\begin{array}{c}
\psi_1 \\ \psi_2
\end{array}
\right) 
= 
 \left(
\begin{array}{c}
\psi_2 \\ \psi_1
\end{array}
\right).  
\label{P}
\ee
If $\psi_1$ and $\psi_2$ are interpreted as the mode amplitudes in two parallel waveguides,
this operator performs the spatial reflection in the  direction perpendicular to the cores. 
On the other hand, the $\mathcal T$ operator represents the effect of the time inversion on the complex amplitudes:
${\mathcal T} \psi_n(\tau)= \psi_n^*(-\tau)$, $n=1,2$.

Like the original equations \eqref{soft} for two anharmonic oscillators, 
their amplitude system \eqref{sys} 
is Hamiltonian.
 It can be written as
\[
i   {\dot \psi_1}=-\frac{\partial {\mathcal H}}{\partial {\psi_2^*}},\qquad
 i  {\dot \psi_2}=-\frac{\partial {\mathcal H}}{\partial {\psi_1^*}},
\]
where the Hamilton function
\begin{align}
{\mathcal H}=(|\psi_1|^2+|\psi_2|^2)(1+\psi_1^*\psi_2+\psi_1\psi_2^*)   \nonumber
\\ + i \gamma (\psi_1\psi_2^*-\psi_1^*\psi_2).
\label{Hcali}
\end{align}

\section{Exact linearisation of the dimer}
\label{Sec4} 

The trivial fixed point of the dynamical system \eqref{sys} is $\psi_1=\psi_2=0$.
In the region  $\gamma >1$,
the fixed point is  unstable (a saddle). 
 With reference to small initial conditions,
 for which the nonlinear terms are negligible,
 it is common to say that 
 the $\mathcal{PT}$-symmetry is spontaneously broken here.
 The term means to indicate that generic small perturbations 
 experience exponential growth. 
   In the region $\gamma \leq 1$, the  trivial fixed point is stable
(a centre) and the symmetry is said to be unbroken.
Small perturbations remain small as $\tau \to \infty$.


Large initial conditions in a
  nonlinear Schr\"odinger dimer may lead to an
  exponential blowup --- similar to the blowup in the symmetry-broken linear dimer ---
  but also may  give rise to stable 
 periodic, quasiperiodic 
 or chaotic orbits \cite{Dima_dimer,Susanto,ABRF,BJF}.
 As a result, 
 the $\mathcal{PT}$-symmetry breaking transition may be 
softened by the nonlinearity.
In order to understand details of the
phase transition in systems modelled by
the
$\mathcal{PT}$-symmetric dimer \eqref{sys},
 we obtain its complete solution here.

Introducing the Stokes variables
\begin{align*}
X= 2(|\psi_1|^2-|\psi_2|^2), \quad
 Y= 2i(\psi_1^*  \psi_2     -\psi_1\psi_2^*),  \end{align*}
 and 
 \[
Z= 2(\psi_1^*\psi_2
+
\psi_1\psi_2^*),
\]
equations \eqref{sys} can be written as a three-dimensional dynamical system:
\begin{subequations}
\label{3D}
\begin{align}
{\dot X}= - 2 \gamma r+(2+Z)Y,   \label{B1}  \\
{\dot Y}= -(2+ Z)X,   \label{B2} \\
{\dot Z}=0.  \label{B3} 
\end{align}
\end{subequations}
Here $r$ is the length of the vector ${\vec r}=(X,Y,Z)$:
\be
r= \sqrt{X^2+Y^2+Z^2}= 2(|\psi_1|^2+|\psi_2|^2).
\label{length}
\ee
The length satisfies
\be
{\dot r}= - 2  \gamma X. 
\label{B4}
\ee

Eq.\eqref{B3} implies that  $Z$ is a conserved
quantity, and  so all trajectories lie on the horizontal planes $Z=const$.
Another conserved quantity is 
\[
\mathcal{H}= \frac{r}{2} \left( 1+ \frac{Z}{2} \right) - \frac{\gamma}{2} Y;
\]
 this is nothing but the Hamilton function \eqref{Hcali}.
The existence of two integrals of motion establishes the complete 
integrability of
the system \eqref{3D} and  the  dimer \eqref{sys}.

We can treat $r$ as an independent variable and add the evolution equation
\eqref{B4} to the system \eqref{3D}. The four-dimensional system \eqref{3D},\eqref{B4}
has one more integral of motion, $I=X^2+Y^2+Z^2-r^2$; solutions of the three-dimensional
system \eqref{3D} are selected by  considering the invariant manifold $I=0$
and restricting to $r \geq 0$. 
The advantage of the four-dimensional formulation is that it
reveals the hidden linearity of the system \eqref{3D}.

Eliminating $Y$ and $r$ from \eqref{B1}, \eqref{B2} and \eqref{B4}, we obtain an
equation of harmonic oscillator or inverted oscillator, 
\be
{\ddot X} + \nu^2 X=0,
\nonumber   
\ee
depending on whether 
\[
\nu^2= (2+Z)^2 -4\gamma^2
\]
is positive or negative.
Once the  two-parameter family of solutions 
for $X(\tau)$ has been written down, 
Eqs.\eqref{B2} and \eqref{B4} can be used to recover the corresponding $Y(\tau)$.

\begin{figure}
\begin{center}
    \includegraphics[width=\linewidth] {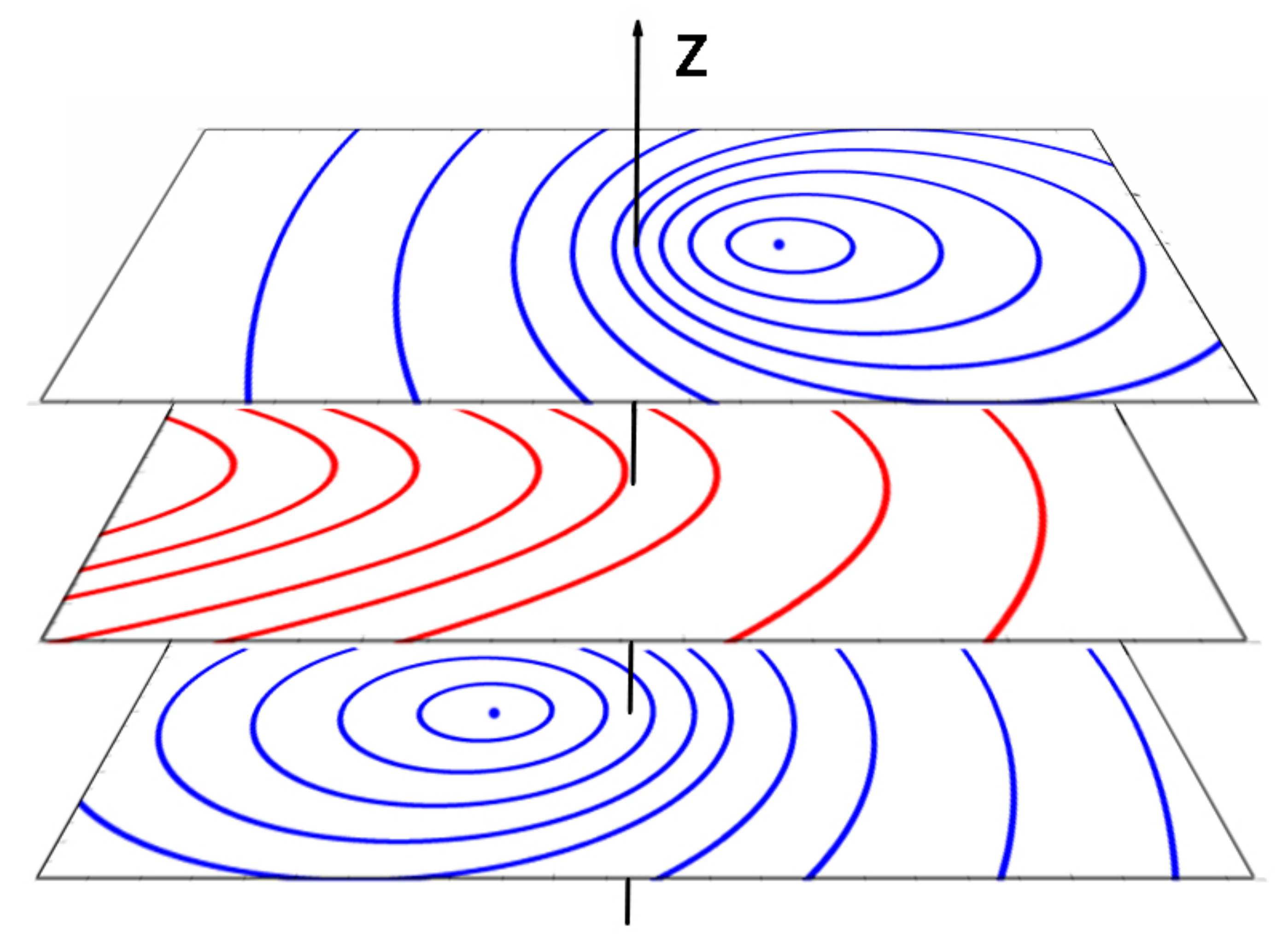}     
   \end{center}
 \caption{The phase space of the system \eqref{3D}.
 All trajectories lie in the horizontal planes $Z=const$.
 The planes with $|Z+2| > 2\gamma$ harbour only periodic orbits
 (solid/blue curves). In contrast, 
 all motions 
 found in the gap  $|Z+2| \leq 2\gamma$ are unbounded
 (dashed/red curves).
 \label{three_D}
  }
\end{figure}

Assume, first, that $(2+Z)^2  >4\gamma^2$, that is, consider 
$Z$ lying below $-2(\gamma+1)$ or above 
$2(\gamma -1)$.
The general solution of the system \eqref{3D} in this case  is
\begin{align}
X= \rho_0 \cos \phi,  \nonumber  \\
Y=Y_0-\frac{2+Z}{\nu} \rho_0 \sin \phi,
\label{XY}
\end{align}
where $\phi=  \nu (\tau -\tau_0)$,
\be
Y_0= \left\{
\begin{array}{rr}
\frac{2\gamma}{\nu} \sqrt{\rho_0^2+Z^2}, & Z>2(\gamma-1); \\ 
\\
-\frac{2\gamma}{\nu} \sqrt{\rho_0^2+Z^2}, & Z<-2(\gamma+1),
\end{array} 
\right.
\label{Y0} 
\ee
and
 $\rho_0>0$, $\tau_0$ are arbitrary constants of integration.
Thus, each horizontal plane with $Z>2(\gamma -1)$ or $Z<-2(\gamma+1)$ hosts a 
family of nested ellipses
\be
X^2+\left( \frac{\nu}{2+Z} \right)^2 (Y-Y_0)^2=\rho_0^2.
\label{ell}
\ee
(See Fig.\ref{three_D}.)
The length of the ${\vec r}$-vector
 remains finite as the imaginary particle moves around the ellipse:
\[
r(\tau)= \frac{2+Z}{2 \gamma}Y_0 - \frac{2 \gamma}{\nu} \rho_0 \sin \phi.
\]

In contrast, all motions corresponding to $Z$  between $-2(\gamma+1)$ and $2 (\gamma-1)$, 
are unbounded:
\begin{align*}
X=-A \sinh s, \nonumber \\
Y= \frac{2+Z}{\sigma} A \cosh s+ Y_0,
\end{align*}
where $s= \sigma (\tau-\tau_0)$,
\begin{align}
\sigma= \sqrt{ 4\gamma^2-(2+Z)^2}>0,
\nonumber \\
A= \sqrt{Z^2+ \frac{\sigma^2}{4 \gamma^2} Y_0^2}>0, \nonumber
\end{align} 
$Y_0$ can be chosen arbitrarily (positive or negative), 
and $\tau_0$ is also an arbitrary parameter.
The length of the  vector ${\vec r}$ in this case is given by 
\[
r= \frac{2 \gamma}{\sigma} A \cosh s + \frac{2+Z}{2\gamma} Y_0.
\]
The solution blows up: $r \to \infty$
as $\tau \to \pm \infty$. (Fig.\ref{three_D}.)

Once we have an explicit expression for 
the trajectory ${\vec r}(\tau)$, 
the corresponding dimer components 
$\psi_1= \sqrt{P_1} e^{i \theta}$ and $\psi_2= \sqrt{P_2} e^{i(\theta-\theta_0)}$ 
can be easily reconstructed:
\begin{align*}
P_1= \frac{r + X}{4}, \quad P_2=\frac{r-X}{4},    \\
\cos \theta_0= \frac{Z}{\sqrt{Y^2+Z^2}}
\quad
\sin \theta_0= \frac{Y}{\sqrt{Y^2+Z^2}},   
\end{align*}
and
\be
\theta= \frac12 \int \left(r+ Z \frac{Z+2}{r+X} \right) d \tau.
\label{pd}
\ee

Finally, we note that the above  solution of the three-dimensional system is in agreement 
with the classification of the $\psi_{1,2}=0$
fixed point in the beginning of this section.
If $\gamma<1$, the ($Z=0$)-plane lies above the  $2(\gamma-1)$ level. Hence the origin $X=Y=Z=0$ is a centre in the $Z=0$ plane;
all trajectories on this and nearby horizontal planes are ellipses.
In contrast, when  $\gamma>1$, the ($Z=0$)-plane lies below  $2(\gamma-1)$ but above the 
$-2(\gamma+1)$ mark. 
In this case, the origin is 
a saddle in its ``plane of residence". Small initial conditions blow up:
$r \to \infty$ as $\tau \to \infty$.

\section{Stationary regimes of the dimer}
\label{Sec5}

In addition to the trivial fixed point at the origin, the
system \eqref{3D}
has a family of nontrivial fixed points $(X_*, Y_*, Z)$.
There is one nontrivial fixed point lying on each horizontal plane $Z=const$
with $Z$ satisfying
\[
(2+Z)^2 \geq 4 \gamma^2.
\]
The horizontal coordinates of the fixed point result  by setting $\rho_0=0$  in \eqref{XY}
and \eqref{Y0}:
\[
X_*=0, \quad
Y_*= 
\left\{
\begin{array}{rr}
\frac{2\gamma}{\nu} |Z|, & Z>2(\gamma-1); \\ 
\\
-\frac{2\gamma}{\nu} |Z|, & Z<-2(\gamma+1).
\end{array} 
\right.
\]

One readily verifies that
for any $\rho_0$,
the distance of the fixed point to the centre of the corresponding ellipse is shorter than its $Y$-semiaxis:
\[
|Y_*-Y_0|<  \frac{|2+Z|}{\nu} \rho_0.
\]
That is, 
the fixed point is enclosed by the entire family of nested ellipses
\eqref{ell},
see  Fig.\ref{three_D}.  This means  that
the  nontrivial fixed point is always stable (a nonlinear centre).

The fixed point of the three-dimensional system \eqref{3D} 
corresponds to a periodic solution of the dimer \eqref{sys}.
However, since the absolute values of the complex amplitudes
$\psi_1 =\sqrt{P_1} e^{i \omega \tau}$ and 
$\psi_2=\sqrt{P_2} e^{i (\omega \tau - \theta_0)}$
are time-independent, this periodic solution represents a stationary configuration of  the
condensate
and describes  a uniform, nonoscillatory,  beam propagation in the optical coupler.
For this reason we will be referring to this solution as the
{\it stationary regime} of the dimer. 
Despite this physically appealing terminology, one should remember  that mathematically, 
the stationary regime is a periodic solution with 
 the
associated frequency $\omega= {\dot \theta}$.

Note that equation $X_*=0$ implies $P_1=P_2$, that is,
the two waveguides carry equal powers in the stationary regime.
(Equivalently, the two modes of the condensate capture equal 
numbers of particles.)

As for 
 the elliptic orbits of the three-dimensional system \eqref{3D}, these
give rise to quasiperiodic solutions of the dimer. 
The corresponding 
 $P_1(\tau)$ and $P_2(\tau)$ are periodic.
Physically, these represent longitudinal variations of the 
optical beam powers and
periodic oscillations of the
 numbers of particles in the 
condensate.

The frequency $\omega={\dot \theta}$ 
is a physically meaningful characteristic of the stationary regime
--- the propagation constant of the optical beam and the chemical potential in the condensate.
 Eq.\eqref{pd} gives $\omega=\Omega(Z)$, where
 the function $\Omega(Z)$ is defined by
 \be
\Omega(Z)= \mathrm{sign} [Z(Z+2)] \, 
\frac{(Z+2)(Z+1)- 2\gamma^2}{\sqrt{(Z+2)^2-4 \gamma^2}}.
\label{Omega}
\ee

In Appendix \ref{Frequency}, we show that  the
equation $\Omega(Z)=\omega$ 
may have one, two, three,  four, or no real roots $Z_n$ --- depending
on $\omega$ and the value of the parameter $\gamma$.
That is, depending on $\gamma$
and $\omega$, there can be one,
two, three or four different stationary regimes of the dimer [nontrivial fixed points of the system
\eqref{3D}] with the given frequency. (Or there may be none.)

Namely, when $\gamma <1$,  there are four different fixed points  for each $\omega> \Omega_1$,
two stationary regimes with the frequency in the range $\sqrt{1-\gamma^2}< \omega< \Omega_1$,
one fixed point for each $\omega$ between $-\sqrt{1-\gamma^2}$ and  $\sqrt{1-\gamma^2}$,
and no stationary regimes if $\omega< -\sqrt{1-\gamma^2}$.  [See Fig.\ref{omega_mu}(a) in the Appendix.]
On the other hand, when $\gamma>1$, the system \eqref{3D} 
 has four fixed points if $\omega> \Omega_1$,
two points in the range $\Omega_3< \omega<\Omega_1$, and no
stationary regimes
 if $\omega< \Omega_3$.
[See Fig.\ref{omega_mu}(c).]
Here $\Omega_1=\Omega_1(\gamma)$ is given by
equation \eqref{Om1}
 with $\varphi$ and $y$ as in \eqref{Z24},
 and
 $\Omega_3(\gamma)$  by equations \eqref{Om3},\eqref{Z24}.
 
 Finally, the case $\gamma=1$ is degenerate.
  In this case the
equation $\Omega(Z)=\omega$ has three roots 
when $\omega>\Omega_1$
and one root for $0<\omega< \Omega_1$.  [See Fig.\ref{omega_mu}(b).] 
Here $\Omega_1=\frac{1+ \sqrt{3}}{\sqrt{2}}3^\frac34$.

 \begin{widetext}

 \begin{figure}
\begin{center}
\includegraphics[width=1\linewidth] {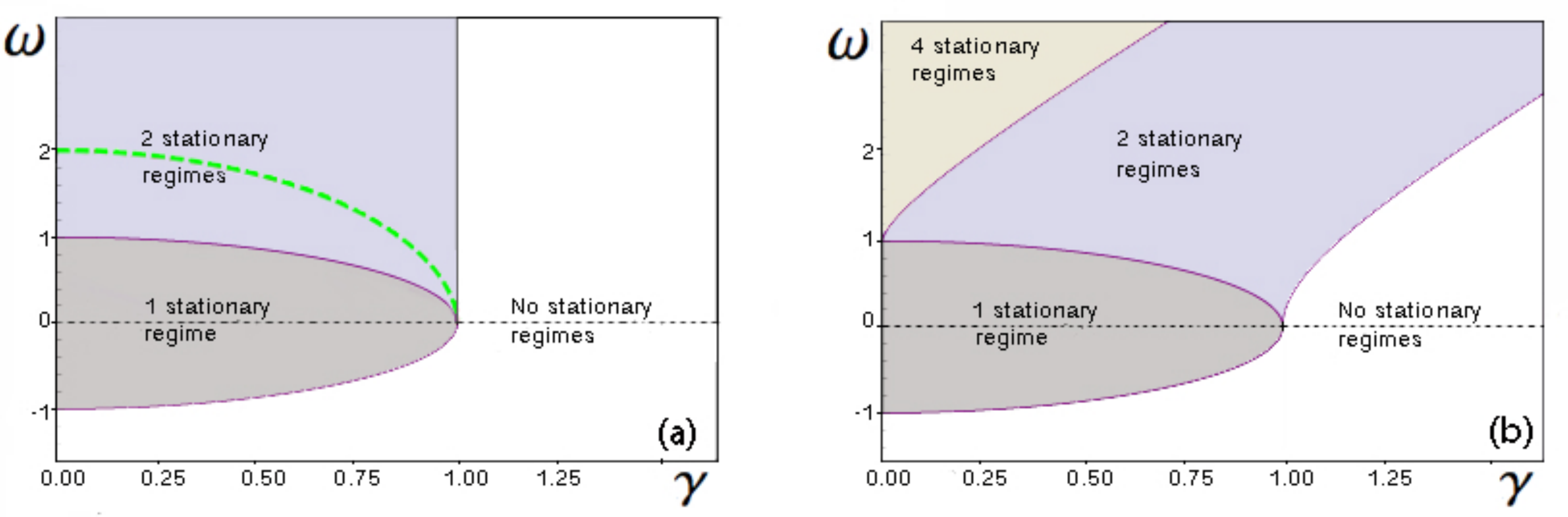} 
    \end{center}
 \caption{ The bifurcation diagram for the ``standard" (linearly-coupled) $\mathcal{PT}$-symmetric dimer (a)
 and its Hamiltonian counterpart (b).
 The $(\gamma, \omega)$ plane is demarcated according to the number of
 co-existing roots of the equation $\Omega(Z)=\omega$. The one-root region
 (shaded grey) in (a) and (b)
    is bounded by $\omega^2+\gamma^2=1$.
 The green dashed  line in (a) is $\omega=2\sqrt{1-\gamma^2}$; above this line the stationary solution \eqref{s1} is unstable. 
 The two-root region (tinted purple) in (b) is bounded by $\omega=\Omega_1(\gamma)$ on the left 
 and $\omega=\Omega_3(\gamma)$ on the right.
 The degeneracy line ($\gamma=1$)  featuring one or three stationary regimes is not marked in (b).
 \label{stab_diagram}
  }
\end{figure}
\end{widetext}

Fig.\ref{stab_diagram}(b)  summarises these conclusions on the $(\gamma, \omega)$ plane. 
The plane has been divided into four domains, according to the number of coexisting 
stationary regimes with the same frequency $\omega$. 
(The degenerate situation along the line $\gamma=1$  has not been indicated.)
The domain boundaries are bifurcation curves of the stationary solutions.

To classify the bifurcations, we note that eigenvalues of the linearisation matrix of Eq.\eqref{sys},
\be
\mathcal L= \left(
\begin{array}{lr}
i \gamma   & 1 \\  1  & -i \gamma
\end{array}
\right),
\label{L}
\ee
are given by $\omega = \pm \sqrt{1-\gamma^2}$. 
Assume $\gamma<1$ is fixed and $\omega$ is increased.
As $\omega$ passes through $-\sqrt{1-\gamma^2}$
[the bottom boundary of the dark-grey region in Fig.\ref{stab_diagram}(b)], one stationary solution bifurcates from 
the eigenvector 
\be
\left( \begin{array}{c} \psi_1 \\ \psi_2 
\end{array}
\right)= 
\left( 
\begin{array}{c}
i \gamma - \sqrt{1-\gamma^2}   \\   1
\end{array}
\right)
\label{EV1}
\ee
 of the matrix \eqref{L}.
 As $\omega$ is increased through $\sqrt{1-\gamma^2}$
 (the top boundary of the dark-grey region), another stationary solution 
 bifurcates from the corresponding eigenvector of $\mathcal L$,
 \be
\left( \begin{array}{c} \psi_1 \\ \psi_2 
\end{array}
\right)= 
\left( 
\begin{array}{c}
1  \\   \sqrt{1-\gamma^2} - i \gamma 
\end{array}
\right).
\label{EV2}
\ee

 As $\omega$ is raised through $\Omega_1$
 [the top boundary of the purple region in Fig.\ref{stab_diagram}(b)], 
 two new stationary solutions appear ``out of the clear blue sky".
 It is important to note that this turning-point bifurcation is not of
 the saddle-centre type, as {\it both} newborn fixed points are centres. 
 The centre-centre folds are not unheard of in the literature; in particular 
 turning points separating two branches of stable solitons 
 were reported in the context of the nonlinear Schr\"odinger equations with
 external potentials \cite{Jianke}.

If $\omega$ is increased for the fixed $\gamma>1$, 
the turning point of this type is encountered twice.
First,  two stable fixed points are born as $\omega$ crosses through $\Omega_3$
[the lower boundary of the purple strip in Fig.\ref{stab_diagram}(b)];
second, two more centres emerge as $\omega$ passes through $\Omega_1$
(the upper boundary of the purple region).

 Another physical  
   characteristic of the stationary regime is
the total  power $P=2P_{1,2}$ carried by the pair of  optical waveguides
--- or,  alternatively, the total number of particles associated with the ground 
and  first excited state in the BEC.
Stationary regimes  have $P=r/2$, with
\be
r={\mathcal R} (Z) \equiv
\frac{|Z(Z+2)|}{\sqrt{(Z+2)^2-4 \gamma^2}}.
\label{rZ}
\ee

\begin{figure}
\begin{center}
    \includegraphics[width=\linewidth] {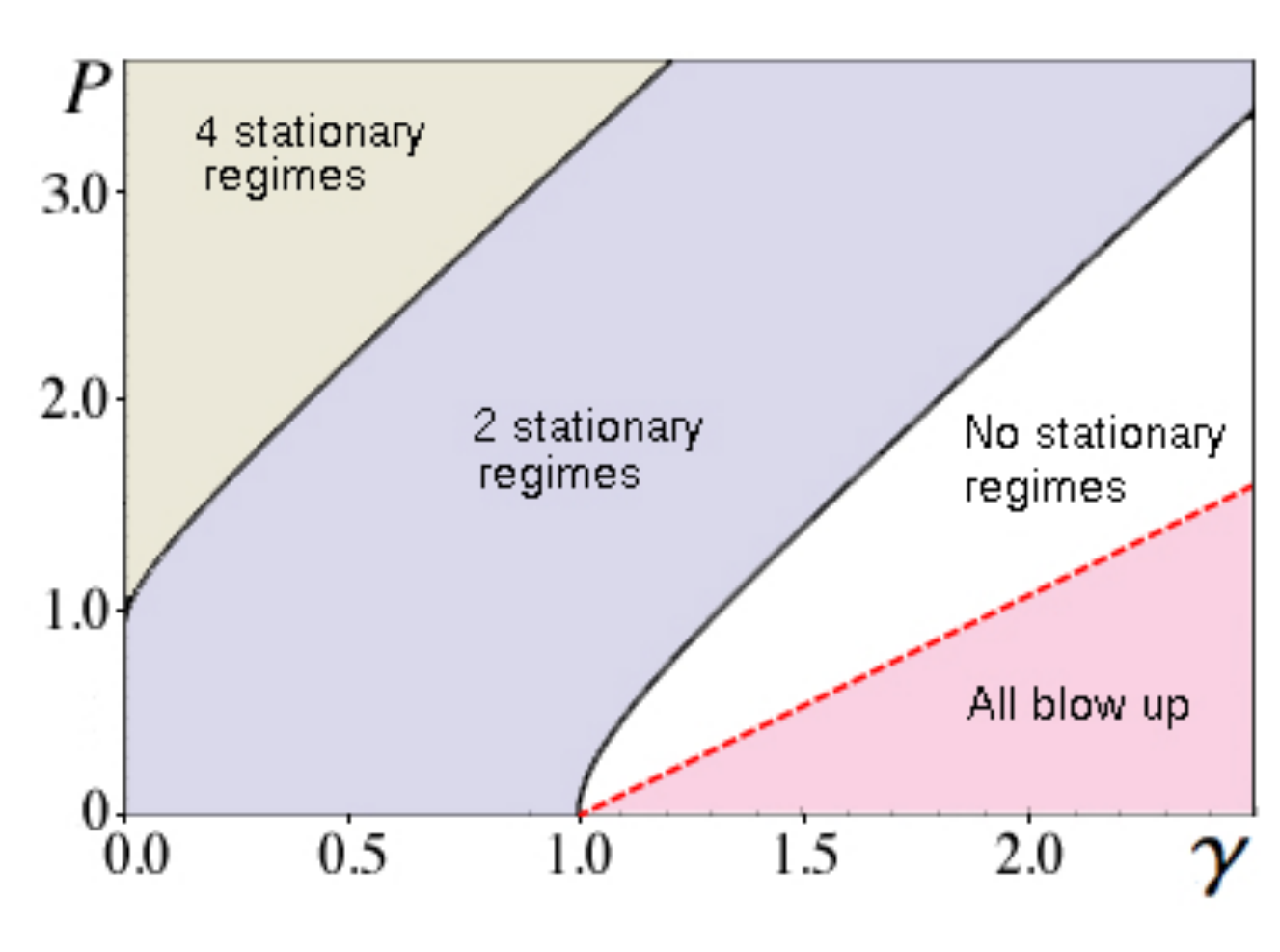} 
    \end{center}
 \caption{The phase diagram for the Hamiltonian dimer \eqref{sys}. 
 The $(\gamma, P)$ plane is divided according to the numbers of coexisting
 stationary regimes carrying the same total power $P$.
 The  boundary between the four- and two-regime phases is given by $P=\frac12 {\mathcal R}_1(\gamma)$.
 As this boundary is crossed from left to right, 
 two of the four stable solutions merge and disappear. The boundary  between the two-regime phase and the phase where no stationary regimes are possible,
 is described by $P=\frac12 {\mathcal R}_3(\gamma)$.
 As this boundary is crossed in the direction of larger $\gamma$,
the two remaining stable stationary solutions merge and disappear.  
In the blank region, $P$ either blows up or performs periodic oscillations.
Dashed  is the line of the $\mathcal{PT}$ symmetry breaking transition, $\gamma=1+P$.
In the pink region below this line, all initial conditions blow up.
 \label{gamma_R}
  }
\end{figure}

In Appendix \ref{power}, we show
 that depending on the value of $r$ and parameter $\gamma$, 
 the equation $\mathcal{R}(Z)=r$ has
two, four, or no roots ${\tilde Z_n}$.
(There is also a degenerate situation where there is one or three roots; see the next paragraph.)
 That is, there can be two or four stationary regimes with the same value of $P$.
 When $\gamma<1$, there are two stationary regimes for each $r<\mathcal{R}_1(\gamma)$
 and four such regimes for $r> \mathcal{R}_1(\gamma)$. [See Fig.\ref{r_mu}(a).]
 When $\gamma>1$, the three-dimensional system \eqref{3D}  has four fixed points for any
 $r> \mathcal{R}_1(\gamma)$,  two such points  for
 $r$ between $\mathcal{R}_3(\gamma)$ and $\mathcal{R}_1(\gamma)$, but no stationary regimes with  $r<\mathcal{R}_3(\gamma)$.
 [See Fig.\ref{r_mu}(c).] These   domains  are
  demarcated in Fig.\ref{gamma_R}.
 
 As with the equation $\Omega(Z)=\omega$, the case $\gamma=1$ is degenerate.
 Here the equation $\mathcal{R}(Z)=r$ has one or three roots, depending on whether
 $r$ is smaller or greater than  $\mathcal{R}_1(1)$ ---
 see Fig.\ref{r_mu}(b).   (Note that this
 degenerate situation is not  delineated in Fig.\ref{gamma_R}.)

\section{Nonlinear  $\mathcal{PT}$-symmetry breaking}
\label{Discussion}

It is interesting to compare the stationary regimes of the Hamiltonian system \eqref{sys}
to those of the ``standard"
 $\mathcal{PT}$-symmetric dimer \cite{Dimer_integrability,Dima_dimer,Susanto,BJF,SXK,LK,Romanian}:
\begin{align}
i {\dot \psi_1} + \psi_2 + |\psi_1|^2 \psi_1 & =-i\gamma \psi_1,   \nonumber \\
i {\dot \psi_2} + \psi_1 + |\psi_2|^2 \psi_2 & = \phantom{-}  i \gamma \psi_2.    \label{R1}
\end{align}
This pair of  coupled equations was used to model a variety of bimodal physical settings, where the
dissipation in one mode is compensated by the energy supply in the other \cite{Kip,DSK,rap_oscillat,LKMG,solitons,ZK,Graefe}.
We note that  the linearly coupled dimer \eqref{R1} governs 
the amplitudes of the $x$ and $y$ oscillations
in the linearly coupled oscillator system 
\begin{align}
{\ddot x} + 2  \eta {\dot x} 
+x + 2 \kappa y 
+  x^3 =  & 0,  \nonumber \\
{\ddot y} - 2 \eta {\dot y}
+ y + 2\kappa x  +  y^3=  & 0.
\label{os}
\end{align}
(See \cite{CKSK}.)
Equations \eqref{os} are reducible to \eqref{R1} in the same way as
 the Hamiltonian oscillator system \eqref{soft} is reducible to the Hamiltonian dimer \eqref{sys}.

The ``standard" dimer \eqref{R1} is integrable \cite{Dimer_integrability,Dima_dimer,Susanto,BJF}.  However no Hamiltonian formulation was
found for this system so far.

When $\gamma$ is smaller than 1, the standard dimer exhibits 
 two stationary regimes of the form $\psi_1=\sqrt{P_1} e^{i \omega \tau}$,
$\psi_2=\sqrt{P_2} e^{i  (\omega \tau -  \theta)}$.  One stationary solution is defined by
\be
P_{1,2} = \omega+\sqrt{1-\gamma^2}, \quad
\theta= \pi - \arcsin \gamma;
\nonumber 
\ee
it bifurcates from the eigenvector \eqref{EV1} as
$\omega$ grows above  $- \sqrt{1-\gamma^2}$
and remains stable for all  $\omega$. 

The second stationary solution  has
\be
P_{1,2} = \omega- \sqrt{1-\gamma^2}, \quad
\theta= \arcsin \gamma;
 \label{s1}
\ee
it  bifurcates from the eigenvector \eqref{EV2} as 
$\omega$ is raised past  $\sqrt{1-\gamma^2}$
and loses stability as $\omega$ is further increased beyond $2\sqrt{1-\gamma^2}$.

The existence and stability domains for the stationary  regimes of Eq.\eqref{R1}  have been demarcated in Fig.\ref{stab_diagram}(a)
--- while Fig.\ref{stab_diagram}(b) lays out a similar bifurcation diagram for the Hamiltonian dimer \eqref{sys}.
The most notable difference between the two panels is that stationary solutions of the
linearly-coupled dimer arise only if $\gamma<1$ \cite{BJF}  whereas the
Hamiltonian dimer admits stable stationary regimes for arbitrarily large values of $\gamma$.

This observation suggests that 
the nonlinearity softens the $\mathcal{PT}$-symmetric 
phase transition in the Hamiltonian dimer. Indeed, the value $\gamma=1$  limits the
stability region of the zero solution in both models --- when $\gamma>1$, small
initial conditions give rise to exponentially growing solutions. 
However when 
these small solutions have grown to become order-one
(or, equivalently, when
order-one initial conditions are considered),
the difference between the two models becomes manifest.
In the standard dimer,  all nonsmall initial conditions  blow up in the same way as the small ones
whereas in its Hamiltonian counterpart, initial conditions with $|\psi_1 \psi_2^*+ \psi_1^* \psi_2 +1|>  \gamma$ lead to bounded trajectories.

To fully appreciate the phenomenon of nonlinear softening, 
it is instructive to introduce the notion of the nonlinear analog of the $\mathcal{PT}$-symmetry breaking.
Consider the optical 
system described by the nonlinear Schr\"odinger dimer \eqref{sys} or \eqref{R1}, and 
denote $P=P_1+P_2$ the total power carried by the pair of waveguides. 
We  say that the  
system suffers the 
$\mathcal{PT}$ symmetry breaking 
transition at the input power level $P$
when $\gamma$
is increased through the point $\gamma_c=\gamma_c(P)$ 
above which all initial conditions with the total power $P$  blow up.


 In the case of the standard dimer, the point of 
the nonlinear $\mathcal{PT}$-symmetry breaking is no different
from the point of the linear phase transition: $\gamma_c(P)=\gamma_c(0)$ \cite{BJF,Susanto}.
In contrast, the Hamiltonian coupler carrying a finite total power $P$
 suffers 
its phase transition for a larger value of the gain-loss coefficient than 
the coupler with the infinitesimal power: $\gamma_c(P) > \gamma_c(0)$.

The exact solvability of our model allows us 
to find the  critical value of $\gamma$  for any $P$.
Indeed,  consider a
ball  $X^2(0)+Y^2(0)+Z^2(0) \leq r_0^2$
of initial conditions
 of the system \eqref{3D}, with the radius 
  $r_0$ satisfying $(2+r_0)^2 > 4 \gamma^2$.
  The ball
   will include initial conditions 
 lying on the horizontal planes $Z=const$ with $(2+Z)^2> 4 \gamma^2$ and therefore
leading to bounded motions.
In contrast, 
a ball of the radius satisfying $2 + r_0 \leq 2 \gamma$
cannot be cut by any horizontal planes with elliptic trajectories.
Since
the total power 
is related to
the length of the vector \eqref{length} 
by $r= 2P$, this simple consideration gives us the critical value of the gain-loss coefficient
for the given total power:
\[
\gamma_c(P)= 1+ P.
\]

\section{Summary and conclusions}
\label{Summary}

\subsection{Summary of results}

We have shown that a pair of coupled nonlinear oscillators, of which
one oscillator has positive and the other one negative damping of equal rate, can form a Hamiltonian system,
Eq.\eqref{soft}.
Small-amplitude oscillations in this  system are described by a  $\mathcal{PT}$-symmetric nonlinear Schr\"odinger dimer
with linear and cubic coupling, Eq.\eqref{sys}.

We have shown that the  dimer \eqref{sys} is completely integrable. 
Unlike   the previously studied linearly-coupled  model  \eqref{R1} (whose Hamiltonian structure has not yet been uncovered),  
the dimer  \eqref{sys} admits a Hamiltonian formulation in terms of the original variables.
Unlike Eq.\eqref{R1},  it is  exactly linearisable
and solvable in elementary functions.

In systems modelled by the Hamiltonian dimer \eqref{sys}, the $\mathcal{PT}$-symmetry breaking 
threshold is determined by the total power:
$\gamma_c=1+P$.
The nonlinearity ``softens" the $\mathcal{PT}$-symmetric phase transition:
no matter how large is $\gamma$, there are  
stable periodic and quasiperiodic  states (with sufficiently high power)
for this  gain-loss rate.

\subsection{Concluding remarks}

Two recent results are worth mentioning 
in the context of the present study.

The first one is due to
Zezyulin and Konotop  \cite{ZK2} who classified  $\mathcal{PT}$-symmetric $N$-component oligomers
with a general cubic nonlinearity.   These authors  prove that   the nonlinearity matrix being
pseudo-Hermitian with respect to the inversion $\mathcal P$  
is sufficient  for the existence of 
  an integral of motion bilinear in ${\vec \psi}$ and ${\vec \psi^*}$.

The dimer \eqref{sys} does share this property;
hence the existence 
of our bilinear conserved quantity $Z= 2(\psi_1\psi_2^*+ \psi_1^*\psi_2)$ follows from the theory of Ref.\cite{ZK2}.
However the second integral of motion, Eq.\eqref{Hcali},  is quartic in the fields so its existence could not
be deduced from their argument.

The second relevant observation belongs to
Pelinovsky, Zezyulin and Konotop  \cite{PZK} who
have demonstrated the integrability of
 the following
 $\mathcal{PT}$-symmetric dimer
 with linear and cubic coupling:
 \begin{align}
i {\dot \psi_1} + \psi_2 +(|\psi_1|^2+|\psi_2|^2) \psi_1 = -i \gamma \psi_1, \nonumber  \\
i{\dot \psi_2} + \psi_1 +(|\psi_1|^2+|\psi_2|^2) \psi_2=  \phantom{-} i \gamma \psi_2.
\label{Manakov}
\end{align}
The nonlinear dynamics exhibited by the system \eqref{Manakov} features
notable differences from that of \eqref{sys}.

\acknowledgements

Discussions with
 G\"unter Wunner 
 and Alexander Yanovski 
are gratefully acknowledged.
We are also  indebted to Valery Shchesnovich for instructive correspondence
and Dmitry Pelinovsky for his critical reading of the manuscript.
One of the authors (MG) thanks Heribert Weigert and Andy Buffler for their hospitality at 
the Centre for Theoretical and Mathematical Physics  and 
Department of Physics
 of UCT.
This project was supported by the NRF of South Africa (grants No 85751, 86991, and 87814).

\appendix
\section{Frequency in stationary regimes}
\label{Frequency} 

In this appendix, we demarcate  frequency ranges
pertaining to the gain-loss coefficient $\gamma$ smaller and
greater than 1. Physically,  the frequency $\omega$ 
represents the propagation constant of the stationary light beam
and chemical potential of the boson condensate.

\begin{figure}
\begin{center}
  \includegraphics[width=\linewidth] {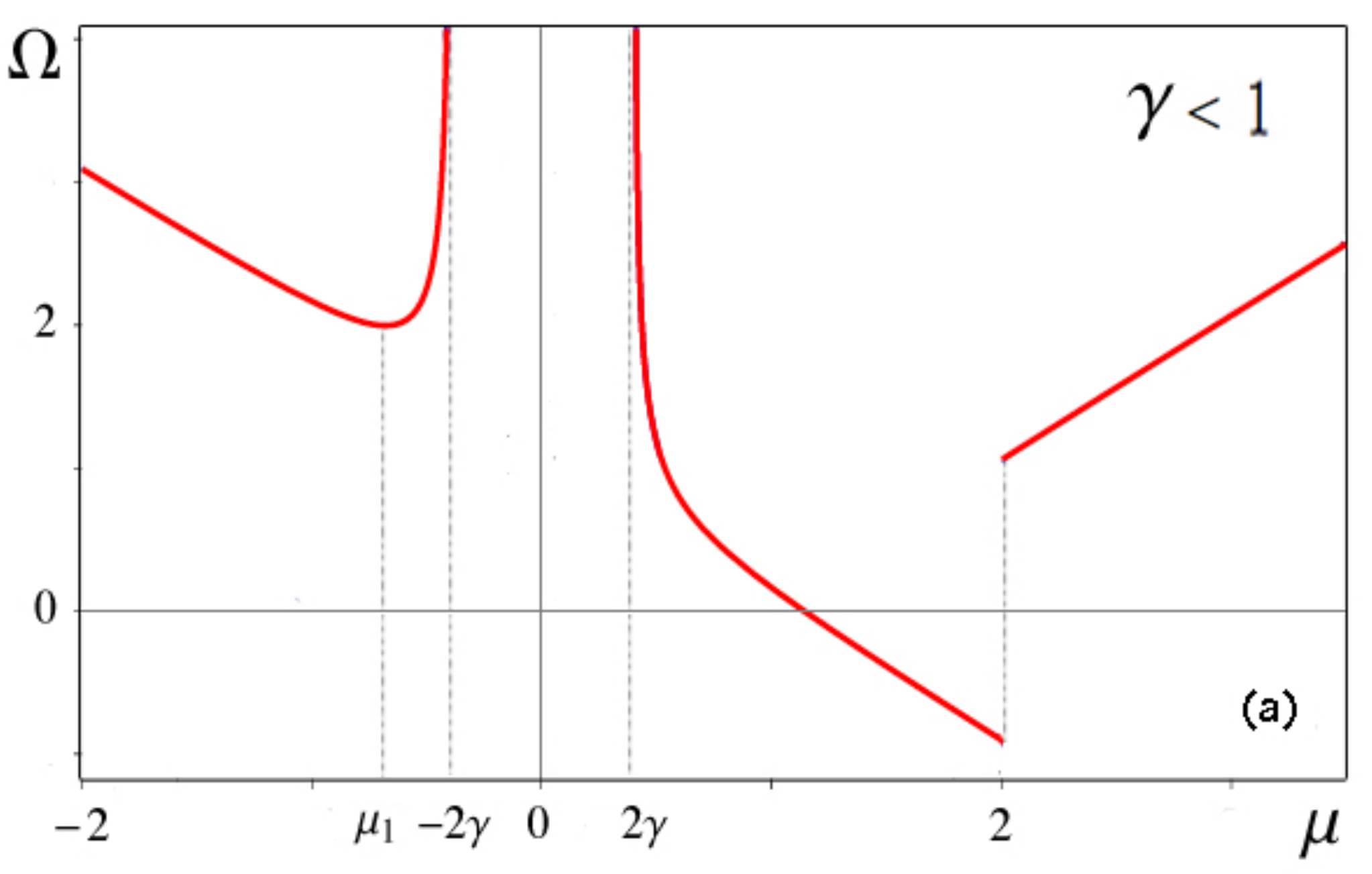}  
   \includegraphics[width=\linewidth] {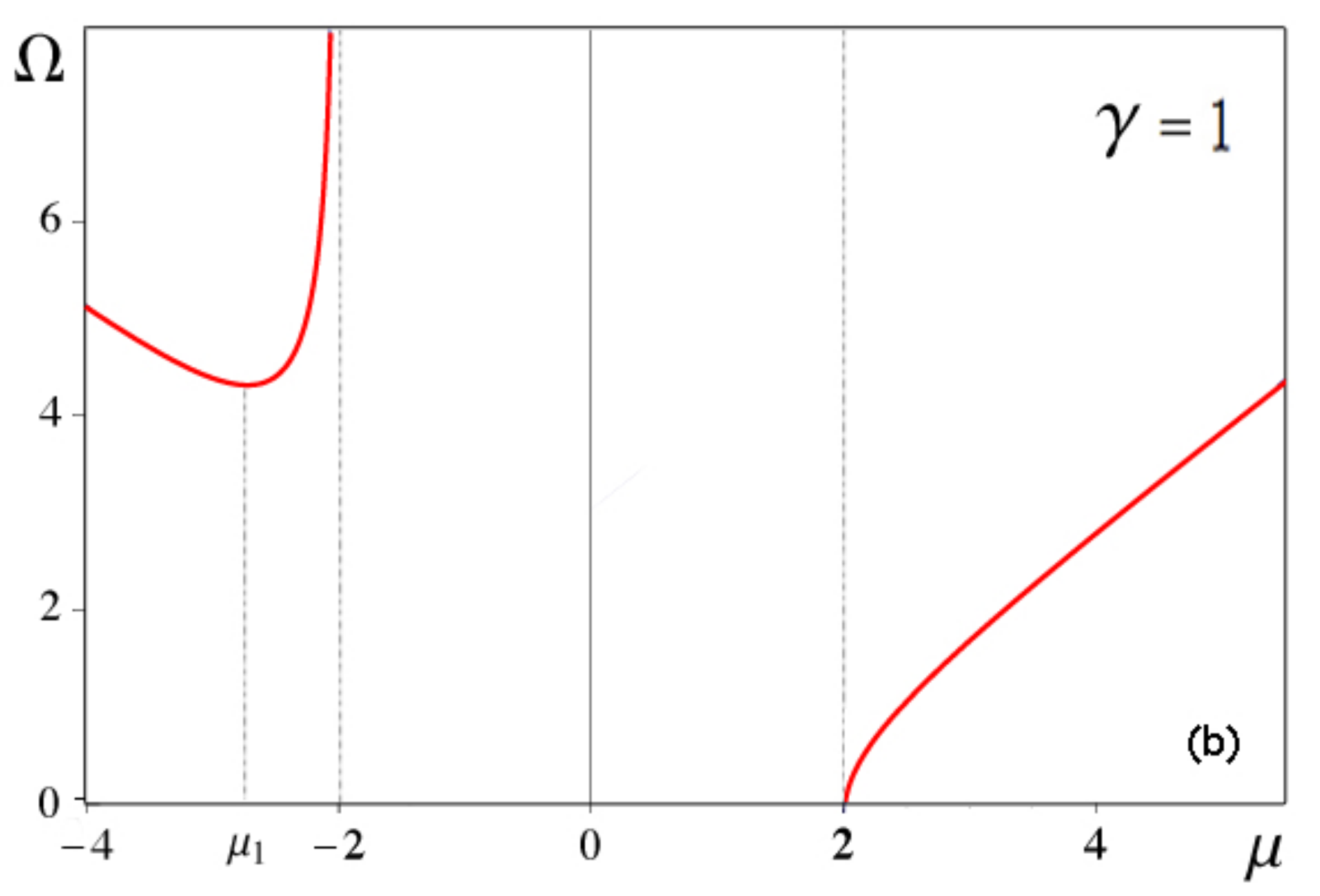}  
     \includegraphics[width=\linewidth] {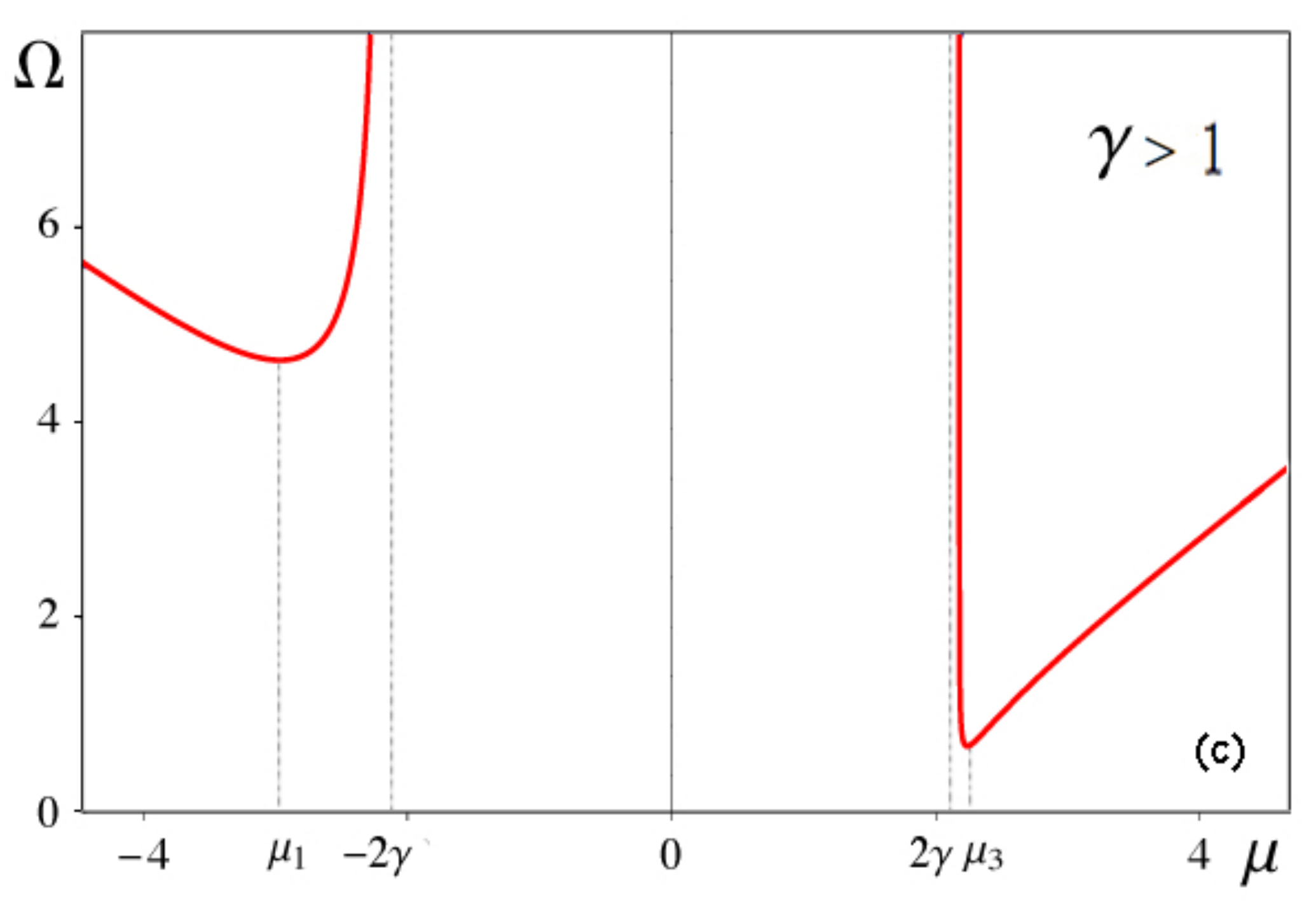}  
   \end{center}
 \caption{Function $\Omega(\mu)$ for $\gamma<1$ (a); $\gamma=1$ (b); and  $\gamma>1$ (c).     
 In (a), $\gamma=0.2$;
 in (c), $\gamma=1.1$.
 \label{omega_mu}
  }
\end{figure}

The frequency function \eqref{Omega} can be written as
 \be
\Omega(\mu)= \mathrm{sign} [\mu (\mu-2)] \, F(\mu), 
\label{Ap1}
\ee
where
\[
F(\mu)=
\frac{\mu(\mu-1)- 2\gamma^2}{\sqrt{\mu^2-4 \gamma^2}}.
\]
Here we have introduced a parameter $\mu=Z+2$,
 which changes from
$-\infty$ to $-2\gamma$ and  from $2\gamma$ to $+\infty$.
The range of frequencies admissible to the periodic solutions of the
system  \eqref{sys} is given by the 
range of the function \eqref{Ap1}.

To determine the range, we need to find the  minima  of $F(\mu)$. The derivative  $dF/d\mu$  vanishes at the points where 
\be
\mu^3- 2 \gamma^2
(3 \mu -2)=0.
\label{Z20}
\ee
Equation \eqref{Z20} has one negative root $\mu_1$, and either 2 complex roots $\mu_2=\mu_3^*$, or 
 2 positive roots $\mu_{2,3}>0$, depending on whether $\gamma<\frac{1}{\sqrt{2}}$ or $\gamma>\frac{1}{\sqrt{2}}$.
 
The negative root is
\be
\mu_1= \left\{
\begin{array}{lr}
-2 \sqrt{2} \gamma \cosh y, 
& 0< \gamma <\frac{1}{\sqrt{2}}; \\
-2 \sqrt{2} \gamma \cos  \varphi,
& \gamma >\frac{1}{\sqrt{2}},
\end{array}
\right.
\nonumber
\ee
where
\be
y=  \frac13  \mathrm{arccosh}  \left(  \frac{1}{\sqrt{2} \gamma}\right),
\quad
\varphi = \frac13  \mathrm{arccos} \left(   \frac{1}{\sqrt{2} \gamma}\right).
\label{Z24}
\ee
As $\mu$ changes from $-\infty$ to $-2\gamma$,
the function $\Omega(\mu)$ decreases from  infinity, reaches a minimum equal to  $\Omega_1>0$,
where 
\be
\Omega_1=\Omega(\mu_1)=
\left\{
\begin{array}{lr} 
\frac{ \gamma+\sqrt{2} \cosh y +2 \gamma \cosh (2y)}
{\sqrt{ \cosh \left( 2 y \right)}}, &  0<\gamma<\frac{1}{\sqrt{2}};   \\
\frac{ \gamma+\sqrt{2} \cos \varphi+2 \gamma \cos (2
\varphi  ) }{\sqrt{ \cos\left( 2 \varphi\right)}}, & \gamma>\frac{1}{\sqrt{2}},
\end{array}
\right.
\label{Om1}
\ee
and then increases back to  positive infinity.
See Fig.\ref{omega_mu} (a-c).

The behaviour of $\Omega(\mu)$ in the region between $2\gamma$ and $+\infty$ 
depends on whether $\gamma$ is smaller or greater than 1.
We first note that since the factor $F(\mu)$ is growing as $\mu \to \infty$,
the rightmost extremum of $F(\mu)$  has to be a minimum.
However since
\[
\frac{d^2 F}{d \mu^2}= 12 \gamma^2 \frac{2\gamma^2-\mu}{(\mu^2-4\gamma^2)^{5/2}}
\]
is negative for $\mu >2\gamma^2$, there cannot be {\it any}  extrema
to the right of $2 \gamma^2$ and
the continuous function $F(\mu)$
has to grow monotonically there.

When $\gamma<1$, the value $2 \gamma^2$ is to the left of $2 \gamma$.
Hence the monotonicity of
the factor  $F(\mu)$ in the region $\mu > 2 \gamma^2$ implies, in particular, that 
 $F(\mu)$ is a monotonically growing function in the whole region $\mu > 2 \gamma$.
As for
$\Omega(\mu)$,
this discontinuous
function
decreases from $+\infty$ to the negative value $-\sqrt{1-\gamma^2} $
as
$\mu$ grows from $2 \gamma$ to 2.
As $\mu$ crosses through $2$, $\Omega$ jumps from $-\sqrt{1-\gamma^2}$ to $\sqrt{1-\gamma^2}$ and then grows to infinity.
See Fig.\ref{omega_mu}(a).

When $\gamma>1$, the function $\Omega(\mu)$ is continuous in its entire domain of definition.
As $\mu$ varies  from $2 \gamma$ to $+\infty$,
$\Omega$ drops from infinity, reaches a minimum at some point $\mu_3$ 
between 
 $2 \gamma$  and $2 \gamma^2$, and grows to infinity as $\mu$ is further increased.
 (There can obviously be no other extrema to the right of $2 \gamma$;
 had there been a maximum there, there would also have to be another minimum
 to the right of it, but this is impossible as the total number of positive extrema is two.)
 See Fig.\ref{omega_mu}(c).
The point of minimum is
 \[
 \mu_3=2\sqrt{2} \gamma  \cos \left( \frac{\pi}{3} - \varphi \right),
 \]
 where 
 $\varphi$ is as in \eqref{Z24}.  
 Denoting $\Omega_3=\Omega(\mu_3)$, we have
 \be
\Omega_3= 
\frac{\gamma
-\sqrt{2} \cos \left( \frac{\pi}{3} - \varphi \right)
+ 2\gamma \cos \left( \frac{2 \pi}{3} - 2 \varphi \right) 
}
{\sqrt{ \cos \left( \frac{ 2 \pi}{3} - 2 \varphi \right) }}.
\label{Om3}
 \ee

We note that  if $\gamma >1$,
$F(\mu) > F(-\mu)$ holds true for any negative $\mu$. 
In particular,  we have $F(\mu_1) > F(-\mu_1)$, where $\mu_1< -2\gamma$ is the point of minimum of the function $F(\mu)$
in the negative semiaxis of $\mu$, and $-\mu_1>2\gamma$ is the symmetrically placed point in the
positive semiaxis. This implies that the function $F(\mu)$ reaches below 
$F(\mu_1)$ in the region  $\mu>2 \gamma$, and hence
$\Omega_3<\Omega_1$ for all $\gamma>1$.

\section{Power in stationary regimes}
\label{power}

\begin{figure}
\begin{center}
 \includegraphics[width=\linewidth] {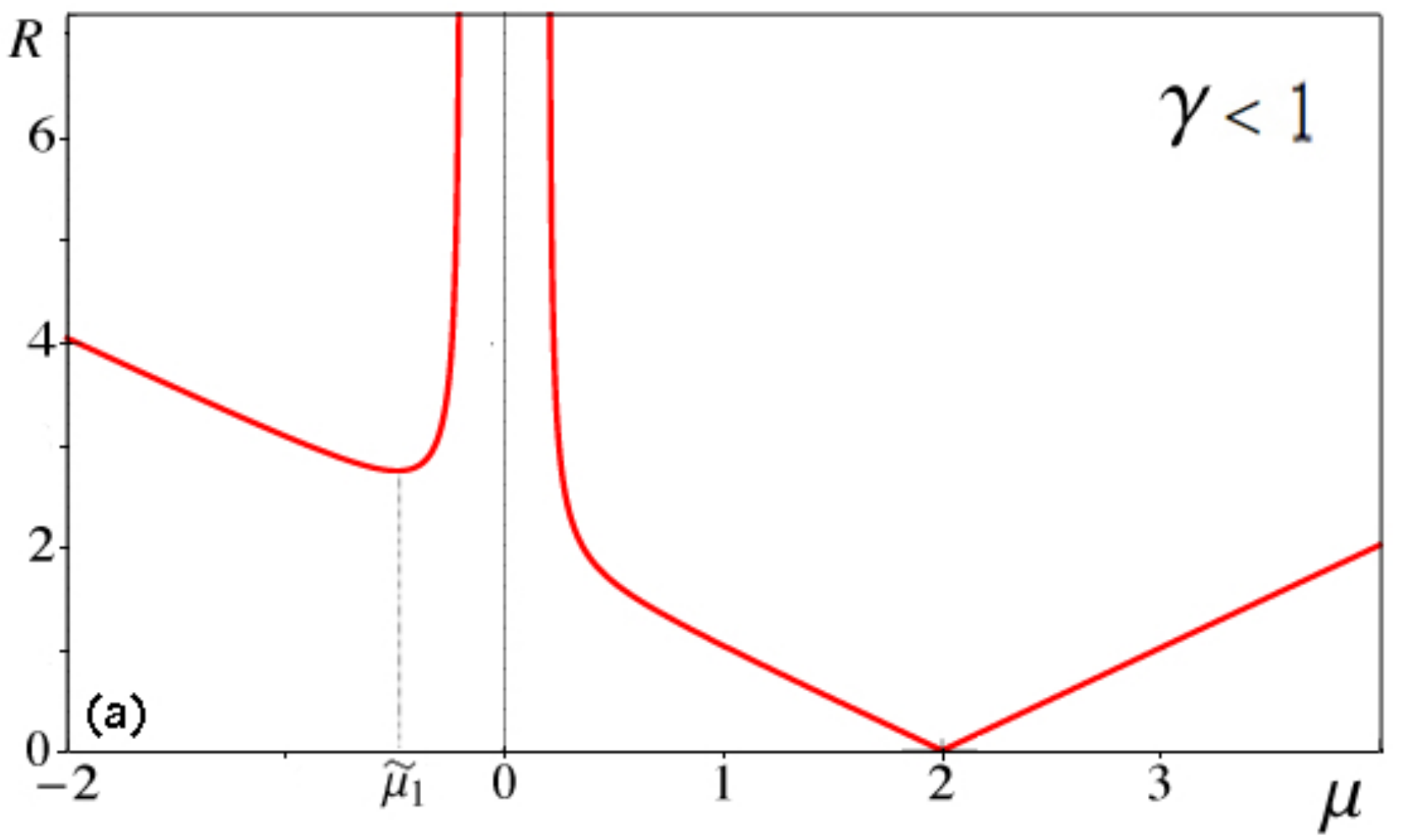}  
  \includegraphics[width=\linewidth] {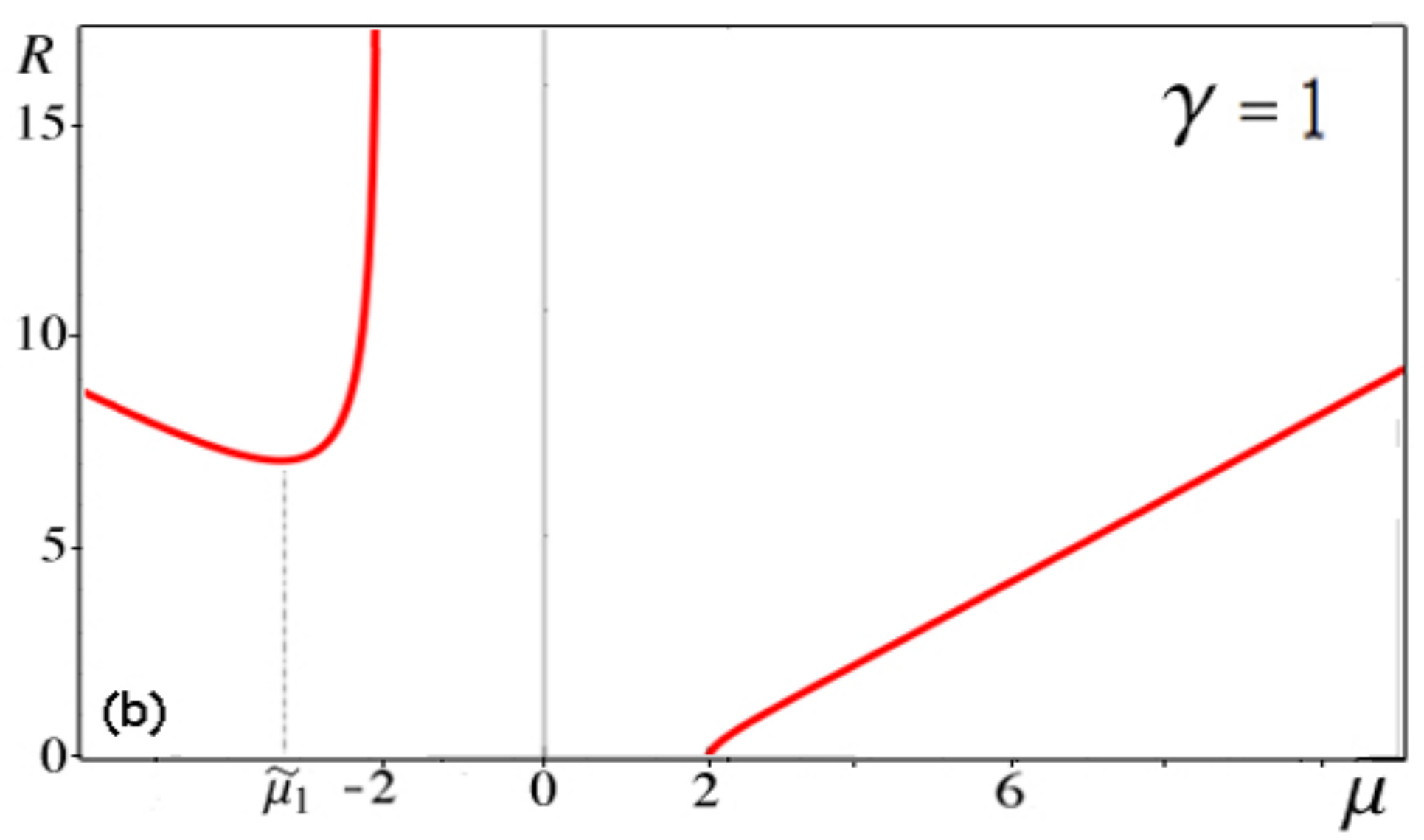} 
    \includegraphics[width=\linewidth] {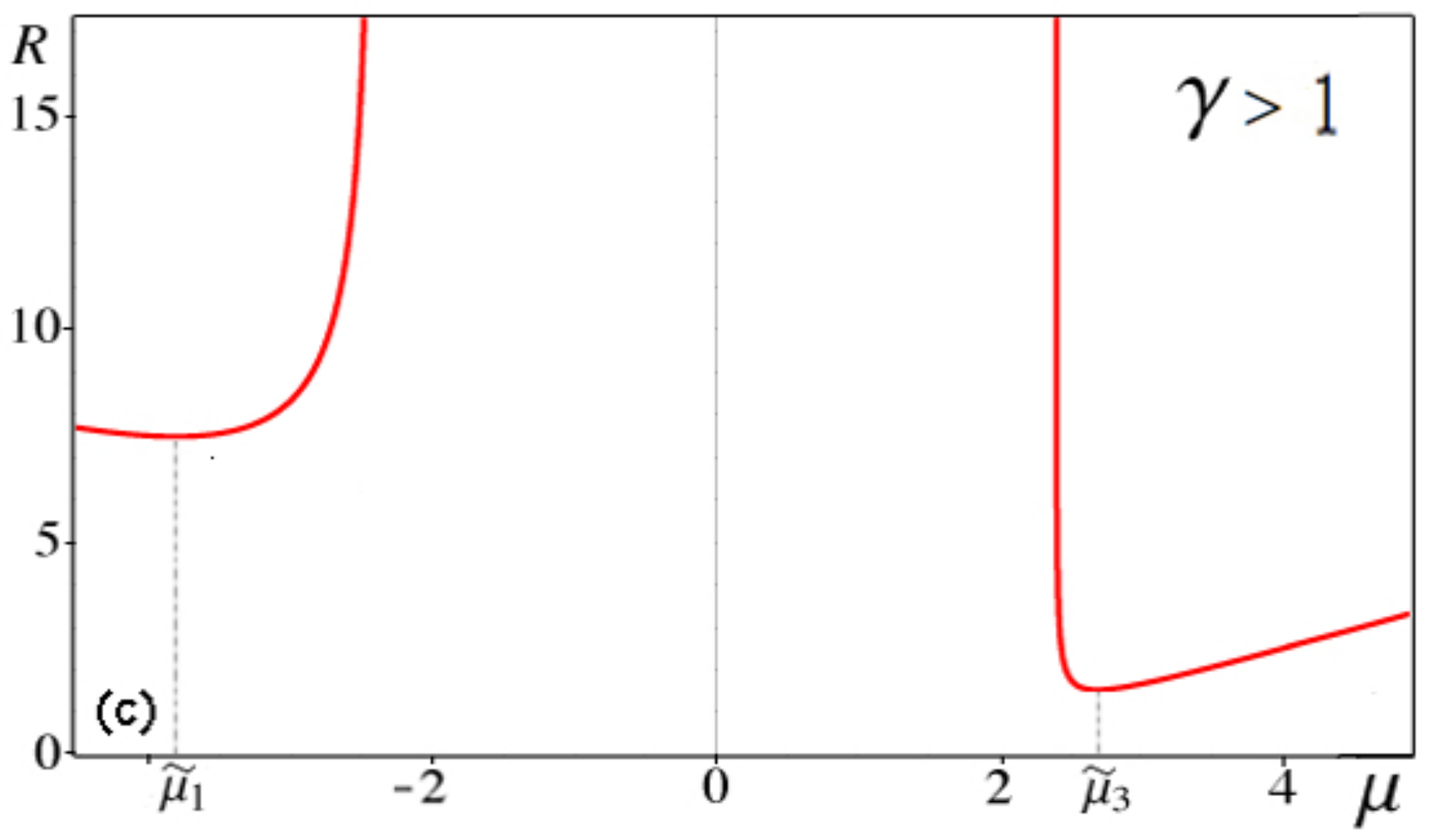} 
   \end{center}
 \caption{Function $R(\mu)$ for $\gamma<1$ (a),  $\gamma=1$ (b),  and  $\gamma>1$ (c).     
 In (a), $\gamma=0.1$;
 in (c), $\gamma=1.1$.
  \label{r_mu}
  }
\end{figure}

The aim of this appendix is to classify 
ranges of the admissible values of power of stationary optical beams
(alternatively,  numbers of particles in the condensate)
described by the dimer \eqref{sys}.

The quantity in question is given by Eq.\eqref{rZ}.
Letting $Z+2=\mu$, we have 
\[
{\mathcal R}(Z)={ R} (\mu) \equiv
\frac{|\mu(\mu-2)|}{\sqrt{\mu^2-4 \gamma^2}}.
\]
The derivative $d {R} /d\mu$ vanishes at the points where
\[
\mathcal{D} (\mu) \equiv \mu^3- 8 \gamma^2(\mu-1)=0.
\]
This equation has one negative root ${\tilde \mu_1}<0$
and either two complex conjugate roots ${\tilde \mu_2}={\tilde \mu_3}^*$
or two positive real roots ${\tilde \mu_{2,3}}>0$ 
depending on whether $\gamma< \sqrt{27/32}$ or $\gamma >\sqrt{27/32}$.

Assume, first, that $\gamma< \sqrt{27/32}$. In this case, the function ${R}(\mu)$ has a smooth minimum
at a negative $\mu={\tilde \mu_1}$ and a cusp at $\mu=2$, with ${R}(2)=0$ (Fig\ref{r_mu}(a)).

In the parameter range $\sqrt{27/32}< \gamma<1$ the function $\mathcal{D}(\mu)$ has a negative
minimum $\mathcal{D}({\tilde \mu_0})<0$, with ${\tilde \mu_0}=\sqrt{8/3} \gamma < 2 \gamma$.
Since $\mathcal{D}(2\gamma)>0$,
the two roots ${\tilde \mu_{2,3}}$
are to the left of $2\gamma$. Therefore, in this case the derivative $dR/d\mu$ does not have any roots in 
the positive part of its domain of existence.
The behaviour of the function 
${R} (\mu)$ coincides with the one shown in Fig.\ref{r_mu}(a).

Finally, it remains to consider the case $\gamma>1$.
 Here we have
 $\mathcal{D} (2), \mathcal{D} (2 \gamma)<0$ while $\mathcal{D}(0)>0$;
 hence  the two positive zeros of $\mathcal{D}(\mu)$
 satisfy $0< {\tilde \mu_2}< 2$ and ${\tilde \mu_3} > 2\gamma$. 
 The point ${\tilde \mu_2}$ is not in the domain of $R(\mu)$
 and cannot be a maximum of this function.
Therefore, the  function $R(\mu)$ only has two minima, at ${\tilde \mu_1}<-2\gamma$  and ${\tilde \mu_3}>2\gamma$.
See Fig.\ref{r_mu}(c).

Expressions for the points of local  minima and the corresponding values of $R(\mu)$ are explicitly available.
The left minimum is at the point
\begin{align*}
{\tilde \mu_1}=  \left\{ 
\begin{array}{lr}
-\sqrt{\frac{32}{3}} \gamma \cosh {\tilde y}, & \gamma < \sqrt{\frac{27}{32}};\\
-\sqrt{\frac{32}{3}} \gamma \cos {\tilde \varphi}, & \gamma> \sqrt{\frac{27}{32}}, 
\end{array}
\right.
\end{align*}
where
\begin{align}
{\tilde y}=  \frac13 \mathrm{arccosh}  \left( \sqrt{\frac{27}{32}} \frac{1}{\gamma} \right),
\quad 
\gamma < \sqrt{\frac{27}{32}};  \nonumber 
\\
{\tilde \varphi}= \frac13  \arccos \left( \sqrt{\frac{27}{32}} \frac{1}{\gamma} \right), \quad \gamma> \sqrt{\frac{27}{32}}.
\label{X1}
\end{align}
The corresponding values  $\mathcal{R}_1=R({\tilde \mu_1})$  are
\begin{align}
\mathcal{R}_1=  
4 \left( \frac23 \right)^{\frac34} \gamma^{\frac12} (\cosh {\tilde y})^{\frac32}  &  \left[  \sqrt{\frac{32}{3}} \gamma \cosh {\tilde y}  +2  \right]^{1/2},   \nonumber \\
\mathcal{R}_1=  4 \left( \frac23 \right)^{\frac34} \gamma^{\frac12} (\cos {\tilde \varphi})^{\frac32}  & \left[   \sqrt{\frac{32}{3}} \gamma \cos {\tilde \varphi}    +2  \right]^{1/2},
\end{align}
for $\gamma$ smaller and larger than $\sqrt{\frac{27}{32}}$, respectively.

The right local minimum (arising only if $\gamma>1$) is at
\[
{\tilde \mu_3}=
\sqrt{\frac{32}{3}} \gamma \cos \left( \frac{\pi}{3} - {\tilde \varphi} \right),
\]
where ${\tilde \varphi}$ is as in \eqref{X1}. 
The corresponding $\mathcal{R}_3=R(\mu_3)$ is given by 
\begin{align}
\mathcal{R}_3=  4 \left( \frac23 \right)^{\frac34} \gamma^{\frac12}   \left[  \cos  \left( \frac{\pi}{3}- {\tilde \varphi} \right)  \right]^{\frac32} 
\nonumber \\
\times    \left[ \sqrt{\frac{32}{3}} \gamma \cos
\left( \frac{\pi}{3}-  {\tilde \varphi} \right) -2    \right]^{1/2}.
\end{align}

Finally,  when $\gamma>1$,
we have $R(\mu)> R(-\mu)$ for any negative $\mu$ in the domain of $R(\mu)$.
By the argument similar to the one produced at the end of Appendix \ref{Frequency}, 
we conclude that ${\mathcal R}_3(\gamma) < {\mathcal R}_1(\gamma)$ for all $\gamma>1$.


\end{document}